\documentclass[12pt]{article}
\usepackage{amsfonts}
\usepackage{amssymb}
\begin{document}
\date{}
\title{\textbf{Noncommutativity from Embedding Techniques}}
\author{\textsc{Saurav Samanta}\thanks{E-mail: saurav@bose.res.in}\\
\\\textit{S.~N.~Bose National Centre for Basic Sciences,}
\\\textit{JD Block, Sector III, Salt Lake, Kolkata-700098, India}}
\maketitle
                                                                                
\begin{quotation}
\noindent \normalsize 
 We apply the embedding method of Batalin-Tyutin for revealing noncommutative structures in the generalized Landau problem. Different types of noncommutativity follow from different gauge choices. This establishes a duality among the distinct algebras. An alternative approach is discussed which yields equivalent results as the embedding method. We also discuss the consequences in the Landau problem for a non constant magnetic field.
\end{quotation}
\noindent{\bf PAC codes:}{03.65.-w, 11.10.Ef}\\

\section{Introduction}

Landau problem is a very good example to understand noncommutativity. Usually people study it by introducing noncommutating coordinates by hand or by introducing noncommutating momenta due to the peculiar structure of canonical momenta which depends on the magnetic field. Various interesting points of this problem has been studied both from theoretical\cite{x} and from phenomenological\cite{y} point of view. However in this paper the embedding method of Batalin-Tyutin\cite{m} is followed. This is done to understand the noncommutativity which arises even in classically in the generalized version Landau problem where a charged particle is subjected to an additional quadratic potential with the usual constant magnetic field. Perhaps this explains the noncommutativity of the problem. 

The paper is organized as follows

In section 2 a brief review of generalized Landau problem is given. In stead of giving a theory on quantum mechanical representation\cite{la} we stressed on the noncommutating Poisson's bracket structure.

Section 3 is dedicated to show the dual nature of different types of noncommutative structure from Batalin-Tyutin\cite{m} point of view. This dual nature is shown to be a consequences of different gauge choices. An interesting discussion for a simpler model is given in \cite{s}.

In section 4 we obtain a mapping between the generalized Landau problem and chiral oscillator for constant parameters. To make this possible both equations of motion and Poisson's brackets are compared. 

Special cases are discussed in section 5 to show that the noncommutativity in the coordinates or in the momentum is actually a dual description of a more general type of noncommutativity. In this way we re-establish the duality of the problem shown in section 2. 
 
In section 6 generalized Landau problem is discussed for non constant magnetic field. Instead of giving a mapping, Poisson's bracket structures are throughly analyzed. It reproduces the result obtained by the symplectic method\cite{b} and also gives a physical realization of the general group theoretical structure discussed in\cite{d}. 

Finally section 7 is for conclusions.
\section{The Model: Generalized Landau Problem}
The classical equations of motion for an electron of charge $e$ moving in the $x_1-x_2$ plane under the influence of a constant perpendicular magnetic field $B$ are,
\begin{equation}
m\ddot x_i=\frac{e}{c}B\epsilon_{ij}\dot x_j.
\end{equation}

The above equations of motion follow from the Lagrangian,
\begin{equation}
L=\frac{m}{2}\dot x_i^2+\frac{e}{2c}B\epsilon_{ij}x_i\dot x_j.
\end{equation}
The canonical momentum is
\begin{equation}
p_i=\frac{\partial L}{\partial \dot x_i}=m\dot x_i-\frac{e}{2c}B\epsilon_{ij}x_j
\end{equation}
while the Hamiltonian is
\begin{equation}
H=\frac{\pi_i^2}{2m}=\frac{1}{2m}\left(p_i+\frac{e}{2c}B\epsilon_{ij}x_j\right)^2.
\end{equation}
Now we generalize the Landau problem by introducing an oscillating potential with spring constant $k$ in the $x_1-x_2$ plane. The equations of motion are\footnote{We have rationalized $e=c=1$.}
\begin{equation}
m\ddot x_i-B\epsilon_{ij}\dot x_j+kx_i=0.
\label{old1}
\end{equation}
It is convenient to express the second order system in its first order form. Furthermore following 't Hooft\cite{a} let the equations of motion of a system $\dot q_i=\{q_i,H\}=f_i(q)$, be a function
of position alone. Then denoting $p_i$ as the conjugate momenta, we note that the Hamiltonian $H=\Sigma_ip_if_i(q)$ does not have a lower bound. A positive-definite function $\rho$, considered as the physical Hamiltonian can be constructed such that $\{\rho,H\}=0$. But this change from the original (unbounded) Hamiltonian $H$ to the bounded positive (semi) definite Hamiltonian $\rho$ leads to a modified algebra that can be obtained as follows\cite{c}:
\begin{equation}
\dot q_i=\{q_i,\rho\}=\{q_i,q_j\}\partial_j\rho(q).
\end{equation}
To reproduce the original set of equations of motion, obviously one should take
\begin{equation}
\{q_i,q_j\}\partial_j\rho(q)=f_i(q),
\end{equation}
leading to a nontrivial algebra of $q_i$, eventually leading to noncommuting structures.

Now the noncommutativity of the generalized Landau problem appears by writing the second order system into a pair of first order equations by doubling the degrees of freedom\cite{c}. Consider the pair of first order equations
\begin{eqnarray}
\dot x_i=\alpha q_i+\beta \epsilon_{ij}x_j
\label{x}\\
\dot q_i=\omega x_i+\lambda \epsilon_{ij}q_j
\label{q}
\end{eqnarray}
which lead to the Landau type equations in both $x_i$ and $q_i$\cite{c},
\begin{equation}
\ddot r_i=(\beta+\lambda)\epsilon_{ij}\dot r_j+(\beta \lambda+\alpha \omega)r_i, \ r_i=x_i,q_i.
\label{old2}
\end{equation}
By identifying,
\begin{eqnarray}
&&\frac{B}{m}=(\beta+\lambda)
\label{rb1}\\
\textrm{and} &&\frac{k}{m}=-(\beta \lambda+\alpha \omega)
\label{rb2}
\end{eqnarray}
eq.\,(\ref{old2}) is regarded as a generic version of eq.\,(\ref{old1}).
Following 't Hooft, a Hamiltonian\cite{c} is constructed,
\begin{equation}
H=(\alpha q_i+\beta \epsilon_{ij}x_j)\pi^x_i+(\omega x_i+\lambda \epsilon_{ij}q_j)\pi^q_i
\label{ham}
\end{equation}
where $(x_i,\pi^x_i)$ and $(q_i,\pi^q_i)$ are canonical pairs. The equations of motion $\dot r_i=\{r_i,H\}$ just yields eqs.\,(\ref{x}) and \,(\ref{q}). As usual, this $H$ is not bounded from bellow. A positive definite $\rho$, commuting with $H$ has to be obtained. This $\rho$ gets identified with the physical Hamiltonian\cite{a}. A natural choice satisfying $\{\rho,H\}=0$ is
\begin{equation}
\rho=\frac{q^2}{2m}+\frac{1}{2}kx^2, \  q^2=q_i^2 \ \textrm{and} \ x^2=x_i^2
\label{19}
\end{equation}
where 
\begin{equation}
\alpha=-\frac{\omega}{km}.
\label{hooft} 
\end{equation}
The corresponding algebra is
\begin{equation}
\{x_i,x_j\}=\frac{\beta}{k} \epsilon_{ij}, \  \{x_i,q_j\}=-\frac{\omega}{k} \delta_{ij}, \  \{q_i,q_j\}=m\lambda \epsilon_{ij}.
\label{chcm}
\end{equation}
This algebra leads to noncommuting structures for both $x_i$ and $q_i$ so that the equations of motion (\ref{x}) and (\ref{q}) can be reproduced from $\dot r_i=\{r_i,\rho\} \ r_i=x_i,q_i$.
Now we are in a position to construct the Lagrangian for this generalized Landau problem. The physical concept behind this construction is given in \cite{fa}.

First a $\Lambda$ matrix is constructed from the basic brackets \,(\ref{chcm}) 
\begin{eqnarray}
\Lambda_{ij}&=&\left[\left\{\Gamma_i,\Gamma_j\right\}\right], \Gamma=(x,q)\nonumber\\&=&\left(\matrix{\frac{\beta}{k}\epsilon_{ij}&-\frac{\omega}{k}\delta_{ij}\cr \frac{\omega}{k}\delta_{ij}&m\lambda\epsilon_{ij}}\right)\nonumber
\end{eqnarray}
Its inverse is
\begin{eqnarray}
\Lambda^{ij}=\frac{k}{\omega^2-mk\beta\lambda}\left(\matrix{mk\lambda\epsilon_{ij}&\omega\delta_{ij}\cr-\omega\delta_{ij}&\beta\epsilon_{ij}}\right).
\label{mat}
\end{eqnarray}
Using eqs. (\ref{hooft}) and (\ref{rb2}) it can be shown $\omega^2-mk\beta\lambda=k^2$.
So one can write eq. (\ref{mat}) as
\begin{eqnarray}
\Lambda^{ij}=\frac{1}{k}\left(\matrix{mk\lambda\epsilon_{ij}&\omega\delta_{ij}\cr-\omega\delta_{ij}&\beta\epsilon_{ij}}\right).
\end{eqnarray}
The Lagrangian is therefore,
\begin{eqnarray}
L&=&\frac{1}{2}\Gamma_i\Lambda^{ij}\dot\Gamma_j-\rho(\Gamma)\nonumber\\
&=&\frac{1}{2k}(mk\lambda\epsilon_{ij}x_i\dot x_j+\beta\epsilon_{ij}q_i\dot q_j+\omega x_i\dot q_i-\omega q_i\dot x_i)-(\frac{1}{2m}q_i^2+\frac{k}{2}x_i^2).
\label{430}
\end{eqnarray}
This was the result obtained in \cite{c} exactly in the same way. The equations of motion derived from above Lagrangian are compatible with \,(\ref{x}) and \,(\ref{q}). Since \,(\ref{x}) and \,(\ref{q}) reproduced (\ref{old2}), this Lagrangian is also compatible with (\ref{old2}) and classically they are equivalent. Of course $x_i$ and $q_i$ satisfy the same equation of motion (\ref{old2}) and hence there is a symmetry between them.
\section{Noncommutativity from Batalin-Tyutin Framework}
In the Batalin-Tyutin\cite{m} formalism the original system which contains the second class constraints is embedded in an enlarged phase space i.e. new auxiliary variables are introduced with the physical degrees of freedom in such a way that the redefined constraints are first class and hence the resulting system is gauge invariant. Then one can construct the first class Hamiltonian from the idea of improved function. A short summary of this theory is given in \cite{s}. In general due to the non-linearity in the second class constraints, the improved function may take the form of infinite series. Such situation has been encountered in \cite{new}. Finally physical degrees of freedom are recovered by imposing gauge conditions. Systems of different structure result from different gauge choices, though they are all gauge equivalent.

The Lagrangian given by eq.(\ref{430}) contains the following constraints
\begin{eqnarray}
&&\Omega^1_i=\pi^x_i+\frac{1}{2k}(mk\lambda\epsilon_{ij}x_j+\omega q_i)\approx 0
\label{con1}\\
&&\Omega^2_i=\pi^q_i+\frac{1}{2k}(\beta\epsilon_{ij}q_j-\omega x_i)\approx 0.
\label{con2}
\end{eqnarray}
The commutator matrix for the above constraints is given by
\begin{eqnarray}
\Omega^{XY}_{ij}&=&\{\Omega^X_i,\Omega^Y_j\};X,Y=1,2\\
&=&\frac{1}{k}\left(\matrix{mk\lambda\epsilon_{ij}&\omega\delta_{ij}\cr-\omega\delta_{ij}&\beta\epsilon_{ij}}\right).
\end{eqnarray}
Since the constraint matrix is nonsingular, according to Dirac's classification\cite{n}
 (\ref{con1}) and (\ref{con2}) are second class constraints. The relevant Dirac brackets are
\begin{eqnarray}
\{x_i,x_j\}^{\star}=\frac{\beta}{k} \epsilon_{ij}, \  \{x_i,q_j\}^{\star}=-\frac{\omega}{k} \delta_{ij}, \  \{q_i,q_j\}^{\star}=m\lambda \epsilon_{ij}
\label{dirac}
\end{eqnarray}
which reproduces (\ref{chcm}). This algebra shows a more general type of noncommutativity than that of \cite{s} where momenta-momenta bracket is zero.

In order to convert the second class constraints (\ref{con1}) and (\ref{con2}) into first class constraints a canonical set of auxiliary variables is introduced
\begin{eqnarray}
\{\phi^X_i,\phi^Y_j\}=\epsilon_{XY}\delta_{ij}; \ X,Y=1,2.
\end{eqnarray}
Now we define the following constraints
\begin{eqnarray}
&&\Psi^1_i=\Omega^1_i+A(\phi^1_i+\epsilon_{ij}\phi^2_j)
\label{cons1}\\
&&\Psi^2_i=\Omega^2_i+C\phi^2_i+D\epsilon_{ij}\phi^1_j
\label{cons2}
\end{eqnarray}
where
\begin{eqnarray}
A=\left(\frac{m\lambda}{2}\right)^{1/2}, \ C=\left(\frac{1}{2m\lambda}\right)^{1/2}\left(1-\frac{\omega}{k}\right), \ D=\left(\frac{1}{2m\lambda}\right)^{1/2}\left(1+\frac{\omega}{k}\right).
\label{sC}
\end{eqnarray}
These values of the coefficients are chosen in such a way that$\{\Psi^X_i,\Psi^Y_j\}=0$
i.e. (\ref{cons1}) and (\ref{cons2}) can be made first class constraints. Also note that $A(C+D)=1$.

To obtain the first class Hamiltonian we begin by constructing the improved variables\cite{s}. Improved variables are first class counterparts of the original variables $x_i$ and $q_i$. These are given by
\begin{eqnarray}
&&\tilde{x_1}=x_1+C\phi^2_1-D\phi^1_2, \  \ \tilde{x_2}=x_2+D\phi^1_1+C\phi^2_2\nonumber\\
&&\tilde{q_1}=q_1+A(\phi^2_2-\phi^1_1), \  \ \tilde{q_2}=q_2-A(\phi^1_2+\phi^2_1)\nonumber
\end{eqnarray}
where $A,C$ and $D$ are given by (\ref{sC}). One can easily check
\begin{eqnarray}
\{\tilde{r_i},\Psi^X_i\}=0; \ \tilde{r_i}=\tilde{x_i},\tilde{q_i}
\end{eqnarray}
so that they are first class indeed. They satisfy the algebra 
\begin{eqnarray}
\{\tilde{x_i},\tilde{x_j}\}=\frac{\beta}{k} \epsilon_{ij}, \  \{\tilde{x_i},\tilde{q_j}\}=-\frac{\omega}{k} \delta_{ij}, \  \{\tilde{q_i},\tilde{q_j}\}=m\lambda \epsilon_{ij}
\end{eqnarray}
which mimics (\ref{dirac}) and is a consequence of a general theorem\cite{m} which states that $\{\tilde{A},\tilde{B}\}= \widetilde{\{A,B\}^{\star}}$.

Any function of the phase space variables can be made first class by the following transformation
\begin{eqnarray}
F(x,q)\rightarrow \tilde{F}(\tilde{x}\tilde{q})=
\left.F(x,q)\right|_{x=\tilde{x},q=\tilde{q}}.
\end{eqnarray}
Hence the first class Hamiltonian is given by,
\begin{eqnarray}
\tilde{H}=\frac{1}{2m}\tilde{q_i}^2+\frac{k}{2}\tilde{x_i}^2.
\end{eqnarray}
It is interesting to note that the equations of motion are form invariant i. e. improved variables satisfy the same equations of motion (\ref{x}) and (\ref{q}). This is just a result of the form invariance of the Hamiltonian and the algebra among the basic variables.

In the enlarged space different gauge conditions can be chosen to show the different types of noncommutative structures. For example in the unitary gauge
\begin{eqnarray}
\Psi^3_i=\phi^1_i\approx 0, \ \Psi^4_i=\phi^2_i\approx 0
\end{eqnarray}
we get back the original physical subspace with the algebra (\ref{dirac}).

Next, we choose gauge condition such that $\{x_i,q_j\}^{\star}=\delta_{ij}$ in which case these variables may be regarded as canonical pairs. In one gauge we obtain noncommuting momenta while in the other, noncommuting coordinates are found. Let us choose the gauge conditions,
\begin{eqnarray}
\Psi^3_i&=&sx_i+q_i-A\phi^1_i+A\sqrt{D/C}\epsilon_{ij}\phi^1_j-A\sqrt{C/D}\phi^2_i+A\epsilon_{ij}\phi^2_j\approx 0\\
\Psi^4_i&=&x_i+(l/2A+\sqrt{CD})\phi^1_i-D\epsilon_{ij}\phi^1_j \nonumber\\
&+&C\phi^2_i+(l/2A-\sqrt{CD})\epsilon_{ij}\phi^2_j+l\epsilon_{ij}x_j\approx 0
\end{eqnarray}
where $A,C$ and $D$ are given by the expressions (\ref{sC}) and $s, l$ will be fixed later.\\
Now the constraint matrix is calculated
\begin{eqnarray}
\Psi^{\mu\nu}_{ij}=\{\Psi^{\mu}_i,\Psi^{\nu}_j\}=\Delta_{\mu\nu}\delta_{ij}+E_{\mu\nu}\epsilon_{ij} \ (\textrm{say}); \ \mu,\nu=1,4.
\label{notation}
\end{eqnarray}
The nonzero matrix elements are given below
\begin{eqnarray}
&&\Delta_{13}=-\sqrt{-\frac{mk\lambda}{\beta}}-s, \ \Delta_{24}=\frac{l\omega}{mk\lambda}-\sqrt{-\frac{\beta}{mk\lambda}}\nonumber\\
&&\Delta_{34}=l\sqrt{-\frac{k}{m\beta\lambda}}, \ E_{44}=-\frac{l^2}{m\lambda}.\nonumber
\end{eqnarray}
So the constraint matrix takes the form
\begin{eqnarray}
\Psi^{\mu\nu}_{ij}=\left(\matrix{0&0&\Delta_{13}\delta_{ij}&0
\cr0&0&0&\Delta_{24}\delta_{ij}
\cr-\Delta_{13}\delta_{ij}&0&0&\Delta_{34}\delta_{ij}
\cr0&-\Delta_{24}\delta_{ij}&-\Delta_{34}\delta_{ij}&E_{44}\epsilon_{ij}
}\right)
\end{eqnarray}
whose inverse is
\begin{eqnarray}
\Psi^{(-1)\mu\nu}_{ij}=\left(\matrix{0&\frac{\Delta_{34}}{\Delta_{13}\Delta{24}}\delta_{ij}&-\frac{1}{\Delta_{13}}\delta_{ij}&0\cr-\frac{\Delta_{34}}{\Delta_{13}\Delta{24}}\delta_{ij}&\frac{E_{44}}{\Delta_{24}^2}\epsilon_{ij}&0&-\frac{1}{\Delta_{24}}\delta_{ij}\cr\frac{1}{\Delta_{13}}\delta_{ij}&0&0&0\cr0&\frac{1}{\Delta_{24}}\delta_{ij}&0&0}\right)
\end{eqnarray}
The relevant Dirac brackets are
\begin{eqnarray}
\{x_i,x_j\}^{\star}=0, \ \{q_i,q_j\}^{\star}=\frac{E_{44}}{\Delta_{24}^2}\epsilon_{ij}, \ \{x_i,q_j\}^{\star}=\frac{\Delta_{34}}{\Delta_{13}\Delta_{24}}\delta_{ij}.
\end{eqnarray}
Moreover if we assign the following values of the coefficients,
\begin{eqnarray}
&&l=\sqrt{-\frac{m\beta}{k\lambda}}\left(\frac{\omega}{k\lambda}-\sqrt{-\frac{m}{\lambda B}}\right)^{-1}\\
&&s=-\sqrt{\frac{kB}{\beta}}-\sqrt{-\frac{mk\lambda}{\beta}}
\end{eqnarray}
we get the algebra
\begin{equation}
\{x_i,x_j\}^{\star}=0, \ \{x_i,q_j\}^{\star}=\delta_{ij}, \ \{q_i,q_j\}^{\star}=B\epsilon_{ij}.
\label{mor1}
\end{equation}
This yields the standard commutative Landau model algebra where the bracket among the momenta gives the magnetic field.

Alternatively, we choose the following gauge constraints
\begin{eqnarray}
\Psi^3_i&=&q_i-A\phi^1_i-(A\sqrt{D/C}+l/2C)\epsilon_{ij}\phi^1_j \nonumber\\
&+&(A\sqrt{C/D}-l/2D)\phi^2_i+A\epsilon_{ij}\phi^2_j+l\epsilon_{ij}q_j\approx 0\\
\Psi^4_i&=&vx_i+q_i-v\sqrt{CD}\phi^1_i-Dv\epsilon_{ij}\phi^1_j+Cv\phi^2_i+v\sqrt{CD}\epsilon_{ij}\phi^2_j\approx 0.
\end{eqnarray}

Using the same notation (\ref{notation}) the expressions of the nonzero matrix elements are written below
\begin{eqnarray}
&&\Delta_{13}=\sqrt{-\frac{mk\lambda}{\beta}}+\frac{lk}{\beta}, \ \Delta_{24}=v\sqrt{-\frac{\beta}{mk\lambda}}-1\nonumber\\
&&\Delta_{34}=-vl\sqrt{-\frac{k}{m\beta\lambda}}, \ E_{33}=-\frac{l^2k}{\beta}.\nonumber
\end{eqnarray}
So the constraint matrix takes the form
\begin{eqnarray}
\Psi^{\mu\nu}_{ij}=\left(\matrix{0&0&\Delta_{13}\delta_{ij}&0
\cr0&0&0&\Delta_{24}\delta_{ij}
\cr-\Delta_{13}\delta_{ij}&0&E_{33}\epsilon_{ij}&\Delta_{34}\delta_{ij}
\cr0&-\Delta_{24}\delta_{ij}&-\Delta_{34}\delta_{ij}&0
}\right)
\end{eqnarray}
whose inverse is
\begin{eqnarray}
\Psi^{(-1)\mu\nu}_{ij}=\left(\matrix{\frac{E_{33}}{\Delta_{13}^2}\epsilon_{ij}&\frac{\Delta_{34}}{\Delta_{13}\Delta{24}}\delta_{ij}&-\frac{1}{\Delta_{13}}\delta_{ij}&0\cr-\frac{\Delta_{34}}{\Delta_{13}\Delta{24}}\delta_{ij}&0&0&-\frac{1}{\Delta_{24}}\delta_{ij}\cr\frac{1}{\Delta_{13}}\delta_{ij}&0&0&0\cr0&\frac{1}{\Delta_{24}}\delta_{ij}&0&0}\right)
\end{eqnarray} 
The Dirac brackets are now given by
\begin{eqnarray}
\{x_i,x_j\}^{\star}=\frac{E_{33}}{\Delta_{13}^2}\epsilon_{ij}, \ \{q_i,q_j\}^{\star}=0, \ \{x_i,q_j\}^{\star}=\frac{\Delta_{34}}{\Delta_{13}\Delta_{24}}\delta_{ij}.
\end{eqnarray}
If we assign the following values of the coefficients,
\begin{eqnarray}
&&v=\left(\sqrt{-\frac{\beta}{mk\lambda}}+\sqrt{\frac{B}{m^2k\lambda}}\right)^{-1}\\
&&l=\sqrt{-\frac{m\beta\lambda}{k}}\left(\sqrt{-\frac{m\beta}{B}}-1\right)^{-1}
\end{eqnarray}
then we get the algebra
\begin{equation}
\{x_i,x_j\}^{\star}=\frac{B}{km}\epsilon_{ij}, \ \{x_i,q_j\}^{\star}=\delta_{ij}, \ \{q_i,q_j\}^{\star}=0.
\label{mor2}
\end{equation}
This is the noncommutative Landau model with usual momenta algebra. We thus conclude that the standard (commutative) and noncommutative Landau models are dual aspects of the same parent model.
\section{An Alternative Approach Based on Doublet Structure }
The original model (\ref{430}) has two sets of variables. It is possible to express this by a doublet of models with each one having a single set of variable. This doublet structure is basically the soldering formalism discussed in various papers. A detail discussion with applications is given in \cite{wot}. Consider the Lagrangians 
\begin{eqnarray}
&&L_{+}=-\frac{1}{2}\epsilon_{ij}z_{i}\dot z_{j}-\frac{\omega_{+}}{2}z_i^2
\label{L+}\\
&&L_{-}=\frac{1}{2}\epsilon_{ij}y_{i}\dot y_{j}-\frac{\omega_{-}}{2}y_i^2
\label{L-}
\end{eqnarray}
where we take $\omega_+$ and $\omega_-$ to be greater than zero.

The equations of motion following from (\ref{L+}) and (\ref{L-}) are
\begin{eqnarray}
&&\dot z_i=\omega_{+}\epsilon_{ij}z_j
\label{z}\\
&&\dot y_i=-\omega_{-}\epsilon_{ij}y_j
\label{y}
\end{eqnarray}
while the brackets are\cite{c}
\begin{eqnarray}
\{z_i,z_j\}=-\{y_i,y_j\}=\epsilon_{ij}.
\label{cm}
\end{eqnarray}

 They represent the motion of one dimensional (chiral) oscillators rotating in the clockwise and anticlockwise directions. Suitable combination of these chiral oscillators leads to a two-dimensional oscillator which has been studied in \cite{ra} in the context of Zeeman effect. Here our motivation is to define two variables $x_i$ and $q_i$ from the chiral oscillator variables $y_i$ and $z_i$ in such a way that $x_i$ and $q_i$ satisfy the correct equations of motion and algebras of the generalized Landau problem.
\subsection{Mapping between the equations of motion}
We make an ansatz
\begin{equation}
x_i=az_i+b\epsilon_{ij}z_j+cy_i+d\epsilon_{ij}y_j.
\label{xi}
\end{equation}
Now using eqs.\,(\ref{x}), \,(\ref{z}) and \,(\ref{y}) one can write $q_i$ in terms of $y_i$ and $z_i$
\begin{equation}
q_i=\frac{1}{\alpha}(\beta-\omega_+)(bz_i-a\epsilon_{ij}z_j)+\frac{1}{\alpha}(\beta+\omega_-)(dy_i-c\epsilon_{ij}y_j).
\label{qi}
\end{equation}
Taking the time derivative of above eq. and using the eqs.\,(\ref{z}) and \,(\ref{y})
\begin{equation}
\dot q_i=\frac{\omega_+}{\alpha}(\beta-\omega_+)(az_i+b\epsilon_{ij}z_j)-\frac{\omega_-}{\alpha}(\beta+\omega_-)(cy_i+d\epsilon_{ij}y_j).
\label{qi1}
\end{equation}
Again using eqs.\,(\ref{xi}) and \,(\ref{qi}) in eq.\,(\ref{q}) we obtain
\begin{equation}
\dot q_i=az_i\{\omega+\frac{\lambda}{\alpha}(\beta-\omega_+)\}+b\epsilon_{ij}z_j\{\omega+\frac{\lambda}{\alpha}(\beta-\omega_+)\}+cy_i\{\omega+\frac{\lambda}{\alpha}(\beta+\omega_-)\}+d\epsilon_{ij}y_j\{\omega+\frac{\lambda}{\alpha}(\beta+\omega_-)\}.
\label{qi2}
\end{equation}
So consistency between \,(\ref{qi1}) and \,(\ref{qi2}) demands
\begin{eqnarray}
&&\beta=-\lambda+(\omega_+-\omega_-)
\label{bl}\\
&&\omega=\frac{1}{\alpha}(\lambda+\omega_-)(\lambda-\omega_+).
\label{1000}
\end{eqnarray}
From the above two equations, using (\ref{rb1}), (\ref{rb2}) and (\ref{hooft}) we can show
\begin{eqnarray}
B=m(\omega_+-\omega_-), \ k=m\omega_+\omega_-.
\label{kB}
\end{eqnarray}
This important result shows that magnetic field appears as the difference whereas the spring constant is is a product of the chiral frequencies.
\subsection{Mapping between the algebra}
Using the definitions of $x_i$ and $q_i$ from eqs.\,(\ref{xi}) and \,(\ref{qi}) we get
\begin{eqnarray}
&&\{x_i,x_j\}=(a^2+b^2-c^2-d^2)\epsilon_{ij}=\frac{\beta}{k} \epsilon_{ij}\\
&&\{q_i,q_j\}=\frac{1}{\alpha^2}\{(\beta-\omega_+)^2(a^2+b^2)-(\beta+\omega_-)^2(c^2+d^2)\}\epsilon_{ij}=m\lambda \epsilon_{ij}\\
&&\{x_i,q_j\}=\frac{1}{\alpha}\{-(\beta-\omega_+)(a^2+b^2)+(\beta+\omega_-)(c^2+d^2)\}\delta_{ij}=-\frac{\omega}{k}\delta_{ij}
\end{eqnarray}
where we have used (\ref{cm}) and consistency with the algebra (\ref{chcm}).

The above three equations are not independent. From the last two equations, using (\ref{bl}) and (\ref{1000}) one can obtain the following relations
\begin{eqnarray}
a^2+b^2=\frac{\omega_+(\omega_+-\lambda)}{k(\omega_++\omega_-)}, \ c^2+d^2=\frac{\omega_-(\omega_-+\lambda)}{k(\omega_++\omega_-)}.
\label{cd}
\end{eqnarray}
These pair of equations give the expressions so that variables of the generalized Landau problem can be defined in terms of the chiral variables with the help of eqs. (\ref{xi}) and \,(\ref{qi}). The interesting point is that the coefficients $a,b,c$ and $d$ are not completely determined. Different choices subject to (\ref{cd}) can be made which exactly reproduce the results for different gauge fixings. 
\section{Special Cases}
In \cite{c}, using the soldering method author obtained different noncommutative structures. Here those results are shown to be special cases of a more general mapping obtained in the previous section.

We note that eqs. (\ref{rb1}), (\ref{rb2}) and (\ref{hooft}) already give severe restrictions on the parameters $\alpha,\beta,\omega$ and $\lambda$. In order to give them specific values we set $\omega=-k$ so that $\{x_i,q_j\}=\delta_{ij}$. Now eq. (\ref{hooft}) implies that this choice of $\omega$ fixes the value of $\alpha$ as $\alpha=\frac{1}{m}$. Using these values of $\omega$ and $\alpha$ we get $\beta\lambda=0$ from eq. (\ref{rb2}). That means either $\beta$ or $\lambda$ is zero. In the following two subsections these situations are studied separately.
\subsection{Case 1}
We consider $\beta=0$ first. Then from eq. (\ref{rb1}) $\lambda=\frac{B}{m}$.
Let us now collect the special values of the parameters mentioned so far
\begin{eqnarray}
\alpha=\frac{1}{m}, \ \beta=0, \ \omega=-k, \ \lambda=\frac{B}{m}
\label{ph1}
\end{eqnarray}
For the parameters (\ref{ph1}), the basic brackets followed from (\ref{chcm}) are given by, 
\begin{equation}
\{x_i,x_j\}=0, \ \{x_i,q_j\}=\delta_{ij}, \ \{q_i,q_j\}=B\epsilon_{ij}.
\label{new}
\end{equation}
This structure is same as (\ref{mor1}) and corresponds to the conventional Landau algebra.

Now to find the connection with the chiral oscillator problem, we take eqs. (\ref{bl}) and (\ref{cd}). From eq. (\ref{bl}) we can choose either $\beta$ or $\lambda$ independently. We make the choice $\beta=0$. This implies that $\lambda$ is fixed by the relation,
\begin{equation}
\lambda=\omega_+-\omega_-.
\label{re1}
\end{equation}
Using the above equation we get from (\ref{cd}),
\begin{equation}
a^2+b^2=c^2+d^2=\frac{\omega_+\omega_-}{k(\omega_++\omega_-)}.
\label{abcd}
\end{equation}
Again as mentioned earlier $a,b,c$ and $d$ are not uniquely determined by the eq. (\ref{abcd}). Different choices can be made from the two parameters class of solutions. One can take the symmetrical combination where $a,b,c$ and $d$ are all equal. But to proceed further we make the following asymmetrical choice
\begin{eqnarray}
&&b=d=0 \ \textrm{and}\\
&&a=c=\left(\frac{\omega_+\omega_-}{k(\omega_++\omega_-)}\right)^{1/2}=\chi(\textrm{say})\nonumber
\end{eqnarray}
so that eqs. (\ref{xi}) and\,(\ref{qi}) imply
\begin{eqnarray}
&&x_i=\chi(z_i+y_i)
\label{43}\\
&&q_i=m\chi\epsilon_{ij}(\omega_+z_j-\omega_-y_j).
\end{eqnarray}
Now using eqs. \,(\ref{cm}) the basic brackets are easy to calculate
\begin{eqnarray}
&&\{x_i,x_j\}=0
\label{sag1}\\
&&\{x_i,q_j\}=\frac{m\omega_+\omega_-}{k}\delta_{ij}
\label{sag2}\\
&&\{q_i,q_j\}=\frac{m\omega_+\omega_-}{k}m(\omega_+-\omega_-)\epsilon_{ij}.
\label{sag3}
\end{eqnarray}
This algebra is compatible with \,(\ref{new}) under the identifications (\ref{kB}). Thus we see, by transforming our second order system to a first order one, by introducing an additional variable the noncommutativity is naturally induced. Again this result is reproduced by superposition of two chiral oscillator, which are also first order system. Since the difference in the chiral frequencies is proportional to the magnetic field the connection of two approaches gets established. 
\subsection{Case 2}
To show the situation where the momenta are commuting we take $\lambda=0$, then from (\ref{rb1}) $\beta=\frac{B}{m}$.
So all the values of parameters are listed below.
\begin{equation}
\alpha=\frac{1}{m}, \ \lambda=0, \ \omega=-k, \ \beta=\frac{B}{m}.
\label{49}
\end{equation}
In this case the basic brackets of the Landau problem following from the algebra \,(\ref{chcm}) are
\begin{equation}
\{x_i,x_j\}=\frac{B}{km}\epsilon_{ij}, \ \{x_i,q_j\}=\delta_{ij}, \ \{q_i,q_j\}=0.
\label{new1}
\end{equation}
Note that algebras given by eqs. (\ref{mor2}) and (\ref{new1}) are structurally equivalent.

Now we have to find the similar situation in chiral oscillator problem. In the previous subsection $\beta$ was taken to be zero in eq. (\ref{bl}). Now to generate commutating momenta $\lambda$ is set to be zero. 
Then we take the following asymmetrical choice of the coefficients from eqs. (\ref{cd})
%from the eqs. (\ref{ss2}).
\begin{eqnarray}
&&b=d=0\nonumber \ \textrm{and}\\
&&a=\frac{\omega_+}{\sqrt{k(\omega_++\omega_-)}}, \ c=\frac{\omega_-}{\sqrt{k(\omega_++\omega_-)}}.\nonumber
\end{eqnarray}
Putting these values of the coefficients in eqs. (\ref{xi}) and (\ref{qi}) we 
observe that $x_i$ and $q_i$ are now defined by the relations
\begin{eqnarray}
&&x_i=az_i+cy_i\\
&&q_i=am\omega_-\epsilon_{ij}z_j-cm\omega_+\epsilon_{ij}y_j.
\label{am}
\end{eqnarray}
Using the algebra (\ref{cm}) it is easy to show that they satisfy the following algebra
\begin{eqnarray}
&&\{x_i,x_j\}=\frac{m(\omega_+-\omega_-)}{mk}\epsilon_{ij}
\label{bl1}\\
&&\{x_i,q_j\}=\frac{m\omega_+\omega_-}{k}\delta_{ij}
\label{bl2}\\
&&\{q_i,q_j\}=0.
\label{bl3}
\end{eqnarray}
We note that above algebra and (\ref{new1}) also match under the same identifications (\ref{kB}).

In section 2 we saw from Batalin-Tyutin extended space framework of generalized Landau problem how the usual (commutative) and noncommutative Landau models were related by gauge transformations. Now we have discussed an alternative approach where the general problem is expressed, through certain parameters, by a doublet structure. The similarity is, different parametric choices correspond to distinct gauge fixings in the extended space approach. 
\section{Analysis for Non-constant Parameters}
In this section an interesting type of non commuting structure is obtained which looks different from (\ref{chcm}). The results are purely classical and any type of quantum mechanical effects are not discussed here. Let us consider a general case where $\alpha,\beta,\omega,\lambda$ of eqs. (\ref{x}) and (\ref{q}) are functions of $x_i(i=1,2)$. Following the same method as discussed in section 2 we can easily show, eqs. (\ref{19}) and (\ref{chcm}) imply correct equations of motion (\ref{x}) and (\ref{q}) under $\dot r_i=\{r_i,\rho\}$,$r_i(=x_i,q_i)$ even for non constant parameters.

We use the Jacobi identity
\begin{eqnarray}
\{x_i,\{q_1,q_2\}\}+\{q_1\{q_2,x_i\}\}+\{q_2,\{x_i,q_1\}\}=0.\nonumber
\end{eqnarray}
This yields
\begin{eqnarray}
m\beta \partial_i\lambda-\frac{\omega}{k} \partial_i\omega=0.
\label{mbeta}
\end{eqnarray}
We use two other Jacobi identities
\begin{eqnarray}
&&\{x_i,\{x_1,q_1\}\}+\{x_1,\{q_1,x_i\}\}+\{q_1,\{x_i,x_1\}\}=0\nonumber\\
&&\{x_i,\{x_2,q_2\}\}+\{x_2,\{q_2,x_i\}\}+\{q_2,\{x_i,x_2\}\}=0\nonumber
\end{eqnarray}
and obtain $\beta\partial_i\omega-\omega\partial_i\beta=0.$
From this equation we get a solution
\begin{eqnarray}
\beta=-\theta\omega,\textrm{ $\theta$ is a constant}.
\label{86}
\end{eqnarray}
Put this expression in \,(\ref{mbeta}) to get
\begin{eqnarray}
\theta\partial_i\lambda+\partial_i\frac{\omega}{km}=0.
\label{-theta}
\end{eqnarray}
At this point we recall eq. (\ref{hooft}) i.e. $\alpha=-\frac{\omega}{km}$. This equation is still valid for non-constant parameters because in calculating $\{\rho,H\}$ from eqs. (\ref{ham}) and (\ref{19}) nontrivial brackets were between $x_i, \pi_i^x$ and between $q_i, \pi_i^q$. So we can write the above eq. (\ref{-theta}) as
\begin{eqnarray}
-\theta\partial_i\lambda+\partial_i\alpha=0.
\label{theta}
\end{eqnarray}
We define a new variable $B$ by the relation
\begin{eqnarray}
\lambda=\alpha B.
\label{gr}
\end{eqnarray}
Now eq. \,(\ref{theta}) can be written in terms of $B$ as,
\begin{eqnarray}
(1-\theta B)\partial_i\alpha=\theta \alpha\partial_iB.
\end{eqnarray}
The above equation gives a solution for $\alpha$,
\begin{eqnarray}
\alpha=\frac{1}{m(1-\theta B)}=\frac{1}{m^{\star}}(\textrm{say}).
\label{91}
\end{eqnarray}
Thus using eqs. (\ref{gr}) and (\ref{91}) $\lambda=\frac{B}{m^{\star}}.$ From eqs. (\ref{hooft}) and (\ref{91}) $\omega=-k\frac{m}{m^{\star}},$ and from (\ref{86}) we get $\beta=k\frac{m}{m^{\star}}\theta.$

In terms of the above new parameters we get the following commutation relations from (\ref{chcm})
\begin{eqnarray}
\{x_i,x_j\}=\frac{m}{m^{\star}}\theta\epsilon_{ij}, \ \{x_i,q_j\}=\frac{m}{m^{\star}}\delta_{ij}, \ \{q_i,q_j\}=\frac{m}{m^{\star}}B\epsilon_{ij}.\nonumber
\end{eqnarray}
This type of algebraic structure for the non-constant parameters has similarity with that of the group theoretical structure analyzed by Souriau\cite{d}. So the algebra thus obtained can be seen as a physical realization of Souriau's method. Various quantum mechanical consequences of these brackets were studied in the reference\cite{b}.
\section{Conclusions}
Let us now emphasize the significance of our work. The generalized Landau problem can be regarded as a constrained Hamiltonian system for which a first order formulation is most natural. In this formulation the number of variables is doubled; moreover second class constraints occur. The Poisson brackets therefore get replaced by the Dirac brackets which are finally elevated to the level of commutators. The Dirac brackets among both sets of dynamical variables lead to noncommuting structures.

Next, we have embeded this second class system in an extended space by introducing new pairs of canonical variables such that the original system is converted into a first class one. This embedded system is therefore considered as a true gauge theory. By choosing the unitary gauge which amounts to setting the new variables to zero, the original second class system is recovered . We have then discussed two particular gauge choices in some details. These choices are done such that, in either case, the two sets of dynamical variables can be regarded as coordinates and their conjugate momenta. However one gauge choice leads to commuting coordinates but noncommuting momenta while the other choice yields commuting momenta but noncommuting coordinates. Since these distinct structures follow from the same master gauge theory, a duality is established between them.

We have also discussed an alternative approach where the original model is regarded as being composed of doublet of models. Different parameterizations of this doublet correspond to different gauge fixings in the embedding approach. In this way complete equivalence between these two formulations is established.
%Batalin-Tyutin extended space framework of generalized Landau problem clearly shows the dual nature of different types of noncommutativity. The usual (commutative) and noncommutative Landau models are related only by gauge transformations.

%An alternative approach is discussed where the general problem is expressed, through certain parameters, by a doublet structure. Different parametric choices correspond to distinct gauge fixings in the extended space approach. 

%The case for non constant parameter is also interesting where Jacobi identities give severe restrictions on the form of various parameters. The results of \cite{b} and \cite{d} are reproduced.
\section*{Acknowledgment}
The author thanks R. Banerjee for suggesting this investigation and S. Ghosh for useful discussions.
 
\end{document}